\documentclass[prl,aps,reprint,superscriptaddress,nofootinbib,longbibliography]{revtex4-2}

\pdfoutput=1

\usepackage[colorlinks=true
,urlcolor=blue
,anchorcolor=blue
,citecolor=blue
,filecolor=blue
,linkcolor=blue
,menucolor=blue
,linktocpage=true
,pdfproducer=medialab
,pdfa=true
]{hyperref}
\usepackage[T1]{fontenc}
\usepackage{fontawesome}
\usepackage{feynmf}
\usepackage{graphicx}
\usepackage{enumitem}
\usepackage{latexsym}
\usepackage{amsfonts}
\usepackage{amssymb}
\usepackage{mathrsfs}
\usepackage{color}
\usepackage{amsmath}
\usepackage[capitalise]{cleveref}
\usepackage{slashed}
\usepackage{dcolumn}
\usepackage{verbatim}
\usepackage{comment}
\usepackage{float}
\usepackage{multirow}
\usepackage{xspace}
\usepackage[normalem]{ulem}
\usepackage{lipsum}
\usepackage[dvipsnames]{xcolor}
\definecolor{nicered}{RGB}{200,37,8} 
\definecolor{niceblue}{RGB}{31.,119.,180}
\definecolor{nicegreen}{RGB}{44,160,44} 
\definecolor{orange}{RGB}{255,127,80}
\definecolor{daveblue}{RGB}{70,157,217}
\definecolor{davegreen}{RGB}{134,190,80}
\definecolor{davered}{RGB}{227,120,101}
\definecolor{rebeccapurple}{RGB}{102,51,153}
\definecolor{nuGreen}{RGB}{43,100,25}
\definecolor{muTeal}{RGB}{60,137,137}
\definecolor{tred}{cmyk}{0, 0.8, 0.62, 0.1}

\def\triumf{TRIUMF, 4004 Wesbrook Mall, Vancouver, BC V6T 2A3, Canada}

\begin{document}

\title{Boosting VBF Reconstruction at Muon Colliders}
\affiliation{\triumf}

\author{Carlos Henrique de Lima}
\email{cdelima@triumf.ca}
\affiliation{\triumf}

\begin{abstract}
Forward muon detection at high-energy muon colliders is crucial for resolving the underlying electroweak process. Detecting these muons is challenging in current detector designs, limited by the shielding required to suppress the beam-induced background. This work proposes using asymmetric beam energies to boost one of the forward muons into the detector acceptance, enhancing the ability to distinguish between $W$- and $Z$-initiated vector boson fusion processes. We demonstrate the capabilities of such an asymmetric collider using VBF Higgs production at 3 and 10 TeV muon colliders with modest boost asymmetries. Asymmetric beam configurations can partially recover the physics potential lost in forward regions when detector coverage is limited.
\end{abstract}

\maketitle

\section{Introduction}

Muon colliders are currently under extensive study as a promising candidate for the next-generation high-energy particle physics experiment~\cite{Delahaye:2019omf,Accettura:2023ked,Adolphsen:2022ibf,Narain:2022qud,P5:2023wyd,AlAli:2021let,Black:2022cth}. They offer a unique opportunity to explore the multi-TeV energy scale, where vector boson fusion and scattering increasingly dominate, allowing the precise study of electroweak physics. While the design of a muon collider is in its early stages and quickly evolving, it is important to consider a wide variety of options in terms of machine parameters to maximize its physics reach.

At high energies, electroweak processes, particularly those mediated by vector boson fusion (VBF)~\cite{Costantini:2020stv}, exhibit significantly forward muons or neutrinos. The ability to identify and reconstruct these forward muons is crucial for disentangling the $W$- and $Z$-initiated processes, thereby enhancing the sensitivity to both Standard Model and beyond-the-Standard Model signatures. Several studies have demonstrated that detecting forward muons significantly improves sensitivity to a broad class of signatures~\cite{Ruhdorfer:2024dgz,Bandyopadhyay:2024gyg,Ruhdorfer:2019utl,Ruhdorfer:2023uea,Li:2024joa,Forslund:2022xjq,Forslund:2023reu,Bandyopadhyay:2024plc,Barducci:2024kig,Frigerio:2024jlh,Li:2025ptq}, including making invisible final states accessible. 

Detecting forward muons is highly challenging in current muon collider designs. Most of the limitation comes from the shielding necessary to handle the beam-induced background (BIB) at these colliders~\cite{InternationalMuonCollider:2024jyv}. In-flight decay of the muon results in a secondary particle cloud that propagates alongside the beam. The current shield designs~\cite{Bartosik:2898358,MAIA:2025hzm,Calzolari:2884558} use conical absorbers in the interaction region that limit the angle acceptances to around 10 degrees from the beam line. Further development of how to effectively suppress the beam-induced background can increase the angular acceptance~\cite{Ally:2022rgk,Bartosik:2019dzq}. Still, realistically, it is very challenging for the angular acceptance to be similar to that in electron or proton colliders. 

In this letter, we motivate an alternative axis of optimization for the experimental design, utilizing asymmetric beam energies to facilitate the measurement of these forward muons. The asymmetric energy boosts the angle of the particle produced in the collision. Forward muons traveling in the opposite direction of the boost can now be inside the detector. While muons traveling in the direction of the boost get further forward.  Because we recover only the muons going in the opposite direction of the beam, the asymmetric beam configuration should be seen as a compromise to push for precision on the electroweak sector, if the shielding and detector technology do not significantly improve. Detecting one muon is often enough~\cite{Han:2020pif} to significantly improve the sensitivity for new physics. It is specifically crucial to identify BSM scenarios that couple differently to $W$ and $Z$ and to suppress SM backgrounds in several BSM searches.

\begin{figure}[b!]
\includegraphics[width=0.9\linewidth]{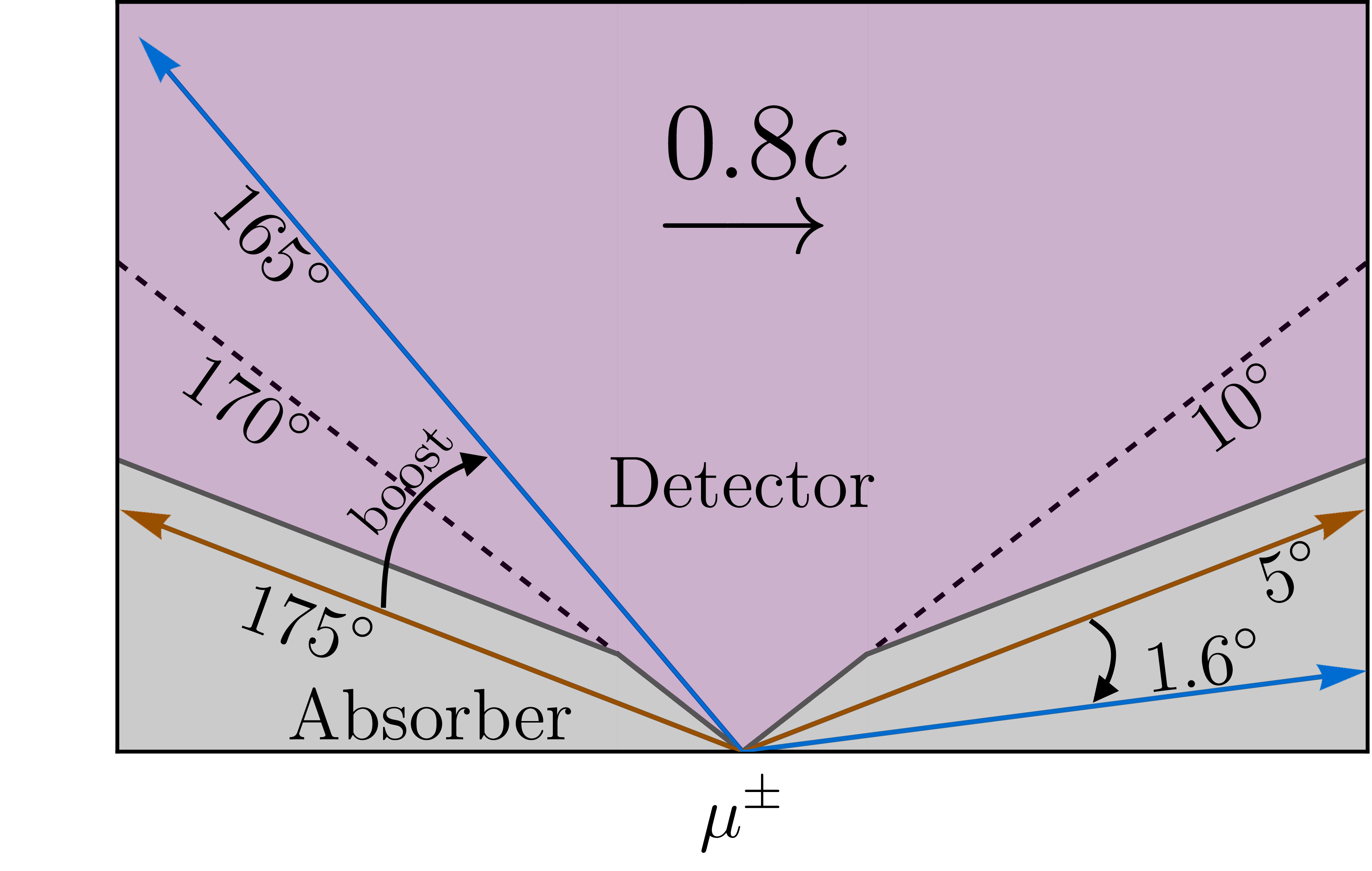}
\caption{ Schematic illustration of the impact of an asymmetric beam configuration ($\beta = 0.8$) on particle angles in the lab frame. A forward muon pair (brown arrows) initially produced at $5^\circ$ relative to the beam axis is Lorentz-boosted such that one muon (blue arrow) enters the detector region. }
\label{fig:angle}
\end{figure}

\begin{figure*}[t!]
\includegraphics[width=0.45\linewidth]{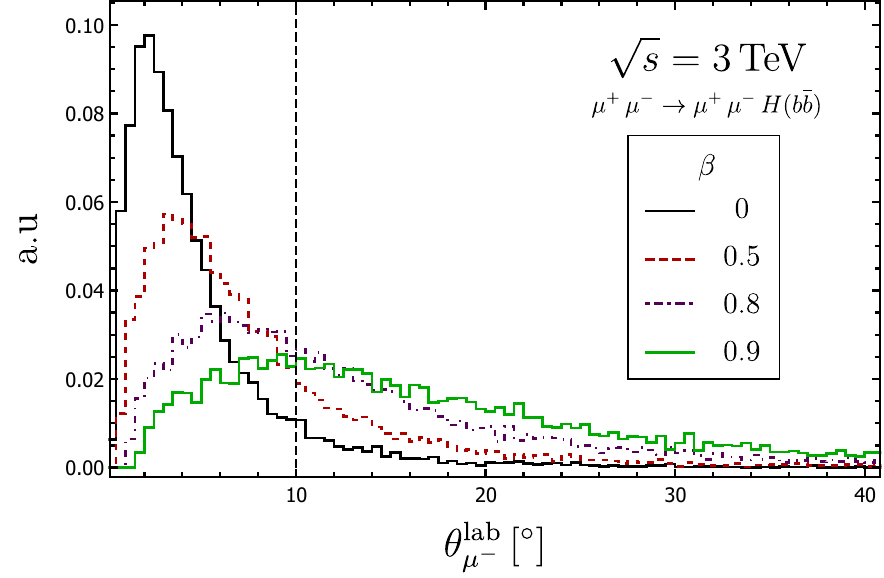}
\includegraphics[width=0.45\linewidth]{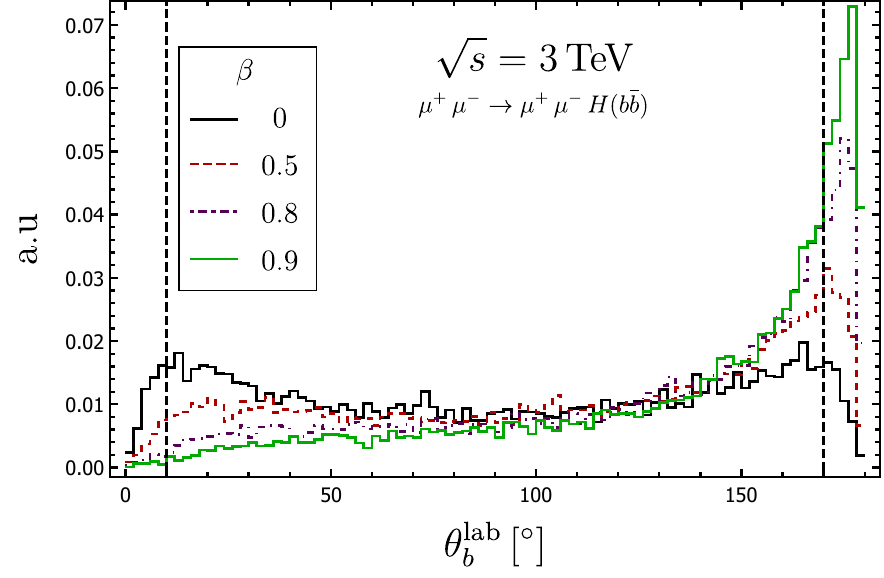}
\caption{\textbf{Left:}  Normalized angular distribution at $\sqrt{s} = 3$ TeV of the muon moving in the opposite direction of the boost. \textbf{Right:} Normalized angular distribution at $\sqrt{s} = 3$ TeV of the $b$-quark from the Higgs decay ($H\rightarrow b\bar{b}$). The different curves show $\beta = 0, 0.5, 0.8, 0.9$. The detector acceptance of 10 degrees from the beamline is shown in dashed black, where most of the muons get lost in the symmetric collider configuration. At the same time, central processes get boosted outside of the acceptance, and the balancing of the two effects is responsible for the better reach for the VBF topology. }
\label{fig:muondis}
\end{figure*}

Asymmetric configurations (which are also asymmetric in lepton type) were explored in~\cite{Lu:2020dkx} and considered for one of the designs of $\mu$TRISTAN~\cite{Hamada:2022mua}. These configurations have a strong physics case to probe lepton number violation~\cite{Kriewald:2024cnt,deLima:2024ohf,Das:2024kyk,Calibbi:2024rcm,Lichtenstein:2023iut}. In this work, we focus on asymmetric muon beams and on their role to enhance sensitivity for VBF topologies. Our results can be easily translated to asymmetric lepton designs. 

Eventually, at multi-TeV collisions, the muons generated from VBF topologies can be so forward that their distribution peaks close to the beamline. The effect of a beam asymmetry using current detector design parameters can be seen in Figure~\ref{fig:angle}. It should be noted that asymmetric configurations give a significant increase in the sensitivity of VBF topologies, but slightly suppresses central processes. The tradeoff is often worth it as the detection of one forward muon is often enough to differentiate W and Z processes, which can be useful for direct BSM searches or to suppress unwanted SM backgrounds.

We demonstrate this idea using VBF Higgs production at a 3 TeV and 10 TeV center-of-mass energy collider with moderate asymmetries with boost $\beta = \frac{E_1-E_2}{E_1+E_2}= [0.5, \, 0.9]$. As the boost increases, the sensitivity for the $Z$-initiated process also increases. This enhances the ability to extract Higgs–gauge boson couplings compared to the symmetric scenario. Even modest asymmetries can improve the discovery reach on electroweak coupled physics without the need for significant progress in detector and shielding design. 

Beyond VBF Higgs production, asymmetric collisions offer a framework to recover physics in otherwise inaccessible regions. Any SM and BSM process with a VBF topology that is generated by $\gamma$ or $Z$ gets benefits from these configurations.  It is important to note that we use VBF Higgs production as a means to highlight the potential to differentiate $W$ and $Z$ VBF topology processes at a muon collider without a forward detector. The ability to discern between $W$ and $Z$ VBF topologies gained from the beam asymmetry is useful to reduce unwanted SM background on diverse BSM searches as Heavy Neutral Leptons~\cite{Kwok:2023dck}, EW axions~\cite{Han:2022mzp,Bao:2022onq,Chigusa:2025otr}, and so on. While it can also be used directly for Higgs and Gauge Bosons precision measurements and other EW-coupled BSM scenarios.

A forward muon detector should be the ultimate goal for maximizing the physics potential of a muon collider. However, it could be that the accelerator challenges are tackled well before the feasibility of those detectors. We propose in this work a different axis of optimization of muon colliders, which maximizes the physics reach for VBF topologies (a major part of the physics program of such colliders), in case no forward detector is available.

\section{Asymmetric Muon Collider}
\label{sec:1}

An asymmetric muon collider consists of two beams with unequal energies, introducing a net boost between the center-of-mass (COM) frame and the laboratory frame. The lab-frame angle $\theta_\text{lab}$ of outgoing particles is related to the COM angle $\theta$ by
\begin{equation}
\cos\theta_{\rm lab}=\frac{\cos\theta+\beta}{1+\beta\cos\theta} \, .
\end{equation}
where $\beta$ is the boost velocity (in units of the speed of light). This transformation compresses forward particles moving aligned with the more energetic beam and expands the acceptance of backward-going particles. An important feature to notice is that the boost effect does not apply to the BIB, since it is an effect of each beam individually.

The tradeoff is that final-state particles emitted in the boost direction are compressed into even smaller angles, potentially exiting the detector coverage as shown in Figure~\ref{fig:angle}. Consequently, reconstructing invisible VBF final states remains challenging, even in asymmetric configurations, as it is not possible to recover both muons.  Nonetheless, the net gain is significant: a single reconstructed forward muon is often sufficient to identify the $Z$-fusion component.

The asymmetric energy configuration should then be seen as a compromise, if the development of the acceleration technology progresses faster than the shielding/detector. In general, if the specific process is mostly backward related to the beam, then there will be a substantial gain in sensitivity. If the process is central, there will be a mild loss in sensitivity, which gets more pronounced as the relative boost increases. 

To visualize this effect, we can use the VBF Higgs production, which will be explored in detail in the next section. In a VBF/VBS topology, the muons are highly forward, but the final state particles generated are mostly central. We can see the effect of including an asymmetric configuration, where we consider the detector geometry between  $10^{\circ} < \theta^{\text{lab}} < 170^{\circ}$. In Figure~\ref{fig:muondis} we have the distributions of the muon going in the opposite direction of the beam, which is highly peaked along the beamline, while the $H(b\bar{b})$ production is approximately flat in the symmetric configuration. As the collider gets more asymmetric, more muons get inside the acceptance, but also more of the $H(b\bar{b})$ get outside of the acceptance. 

\begin{figure}[h!]
\includegraphics[width=0.8\linewidth]{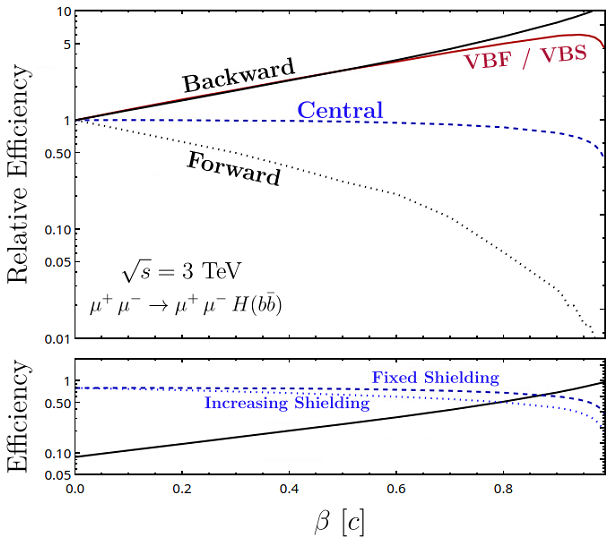}
\caption{Geometric efficiency for VBF Higgs production with a detector coverage of $10^{\circ} < \theta^{\text{lab}} < 170^{\circ}$, shown as a function of the asymmetry boost. The \textbf{Forward} and \textbf{Backward} curves represent the efficiency for detecting the muons along the beamline, while the \textbf{Central} curve shows the efficiency for detecting the $H(b\bar{b})$ $b$-jets. The combined efficiency for the full VBF process is shown in red. The bottom panel illustrates the scenario where the shielding must be extended on the side of the less energetic beam, modeled by a geometric acceptance that degrades linearly with the boost, from  $10^{\circ}$ up to $30^{\circ}$. }
\label{fig:effi}
\end{figure}

We can visualize this directly by comparing the geometric efficiency as a function of the boost, shown in Figure~\ref{fig:effi}. The acceptance of the muon going opposite to the boost gets significantly better, but at the same rate, the acceptance of the muon going along the boost gets significantly worse. If reconstructing one muon is enough (in the next section, we show that it is not only enough but necessary to be able to disentangle Z and W processes), then the asymmetric configuration has a better reach than the symmetric one. At the same time, the effect on central physics is mild, but could still impact the overall reach if there is no corresponding compensation with the luminosity. 

Changing the beam energies can also affect the dynamics of the BIB. The higher energy beam will have a more collimated cloud of particles, while the lower will have a larger spread. It also modifies the energy of the secondary particles that reach the detector after interacting with the nozzle. While we do not simulate the BIB dynamics in this work, we explore what would be the effect if the shielding on the lower energy beam side needs to increase.  To parametrize this, we consider the case where the acceptance gets linearly worse as a function of the boost up to $30^{\circ}$ and show the effect on the efficiency at the bottom of Figure~\ref{fig:effi}. 

Because of the increased spread of BIB in low-energy muon beams, an optimal configuration that maximizes the physics output of asymmetric colliders can be achieved with $\mu e$ colliders, as the one proposed configuration for $\mu$TRISTAN~\cite{Hamada:2022mua}. In this setup, the low-energy beam being $e^{\pm}$ allows for high coverage of the forward region, accessing the particles that get further boosted in that direction. At the same time, the high-energy muon beam allows for a sharper nozzle, potentially increasing the detector range outside of the nozzle. 

It is also important to notice that there may be a potential opportunity cost for these types of colliders. To be able to have an energy asymmetry means that it may be technologically possible to create beams with energy higher than $\sqrt{s}/2$. However, this is not always achievable, depending on the accelerator technology. For example, the designs of $\mu$TRISTAN~\cite{Hamada:2022mua} can only handle $\mu^+$, or plasma-wakefield accelerators~\cite{Lindstrom:2023owp,Foster:2023bmq} that can only accelerate $e^-$.

For this work, we consider the alternative of an asymmetric muon collider not the symmetric configuration with the same energy, but one with significantly higher energy. There may be differences in the instantaneous luminosity\footnote{An asymmetric muon collider can have a mild luminosity loss compared to the symmetric configuration~\cite{Barducci:2023gdc}, scaling with the Lorentz factor as $1/\gamma^2 = 1-\beta^2$. This can be mitigated by increasing the energy spread of the lowest energy beam. The luminosity loss can also be mitigated in colliders of different leptons, with the lower energy beam being $e^{\pm}$. } between these configurations~\cite{Barducci:2023gdc}. To compare these colliders on a level playing field, we consider the same integrated luminosity $\mathcal{L}$ between symmetric and asymmetric configurations. We consider the benchmark collider scenarios in Table~\ref{tab}. 

For this work, we do not simulate BIB of these different beam asymmetries and consider a unique detector design to be able to easily compare the scenarios. Of course, having a full detector simulation that includes a realistic treatment of beam-induced backgrounds would be ideal. However, such simulations depend strongly on the rapidly changing detector and accelerator designs, as well as on the specific experimental setup. In contrast, the advantages we highlight in this work do not rely on these details and remain robust with respect to them. A comprehensive study based on full detector and BIB simulations is therefore beyond the scope of this paper.

\begin{table}[h!]
\begin{tabular}{|c|c|c|c|c|}
\hline
$\sqrt{s}$ & Boost $\beta$ & $(E_1,E_2)$ & Symmetric $\sqrt{s}$ & $\mathcal{L}$ \\ \hline
           & 0.5 &  (2.59, 0.86) TeV   & 5.2 TeV  & \\ \cline{2-4} 
3 TeV      & 0.8 &   (4.50, 0.50) TeV  & 9 TeV   &  1 ab$^{-1}$ \\ \cline{2-4} 
           & 0.9 &  (6.53, 0.34) TeV   & 13 TeV  &  \\ \hline
           & 0.5 &  (8.66, 2.88) TeV   & 17 TeV  &  \\ \cline{2-4} 
10 TeV     & 0.8  &  (15.00, 1.66) TeV  & 30 TeV  &  10 ab$^{-1}$ \\ \cline{2-4} 
           & 0.9  &  (21.79, 1.14) TeV  & 44 TeV  &  \\ \hline
\end{tabular}
\caption{\label{tab}Benchmark collider configurations used in this work. For each value of the center of mass energy $\sqrt{s}$ and boost $\beta$ (in units of the speed of light), we list the energy of each beam and of the symmetric collider with twice the highest beam energy.}
\end{table}

\section{VBF Higgs}
\label{sec:2}

In a muon collider, a single Higgs can be produced by both $W$ and $Z$ fusion, where the $W$ is the dominant production channel, but produces forward neutrinos, and the $Z$ is subdominant and produces forward muons:
\begin{align}
    \mu^+ \, \mu^- &\rightarrow \bar{\nu}_\mu 
    \, \nu_\mu \, H \, , \\
    \mu^+ \,  \mu^- &\rightarrow \mu^+ \, \mu^- \, H \, .
\end{align}

The VBF topology of these processes means that the muons have a forward angle of typically $\theta_\mu \approx m_Z/E_\mu$. As the energy of the beam increases, most of the distribution of the muons gets concentrated below $10$ degrees and thus both processes become indistinguishable as shown in Figure~\ref{fig:muondis}. This, in turn, means that the reach for physics that is produced by $Z$ fusion gets obscured by the $W$ fusion process. As the process has a $p_T$ that is dictated by the vector boson mass, as the energy increases, more forward are the muons which makes detecting them even more difficult.

As the Higgs is produced on-shell, the scale of the process is set by its mass. This eliminates the need to consider complicated electroweak parton distribution functions~\cite{Han:2020uid,*Han:2022laq}. As the asymmetric configuration can only recover one of the forward muons, we consider the visible decay channel, more specifically $H \rightarrow b \bar{b}$. We simulate both processes using \texttt{MadGraph5\_aMC@NLO}~\cite{Alwall:2014hca} interfaced with \texttt{Pythia8.2}~\cite{Sjostrand:2014zea} for showering and hadronization. Jet clustering is done with FastJet~\cite{Cacciari:2011ma} using the Valencia algorithm~\cite{Boronat:2014hva}, with $R=0.5$ and requiring two jets in exclusive mode.

We need to both measure and differentiate the $W$ and $Z$ initiated process, so we consider two separate analyses, following~\cite{Han:2020pif}, which are combined in the end:
\begin{itemize}
    \item Inclusive channel: No muons detected, and thus both $W$ and $Z$ fusion are the signal.
    \item Exclusive $1\mu$ channel: Only events from $Z$ fusion are the signal.
\end{itemize}

\begin{figure*}[t!]
\includegraphics[width=0.32\linewidth]{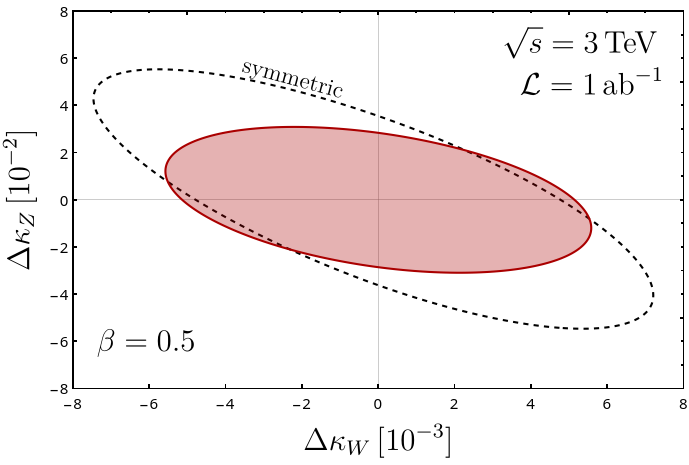}
\includegraphics[width=0.32\linewidth]{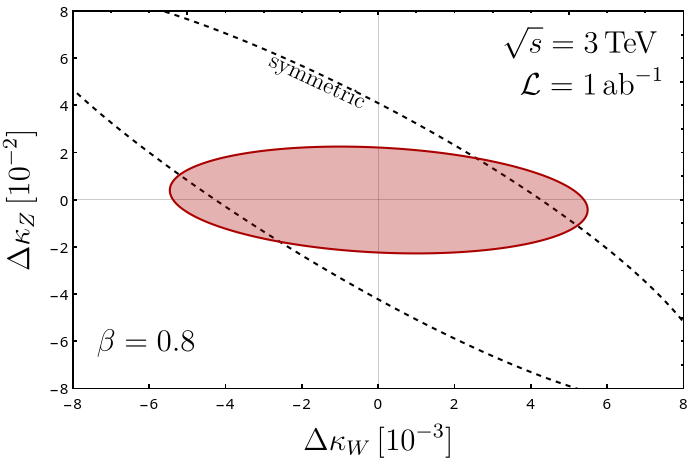}
\includegraphics[width=0.32\linewidth]{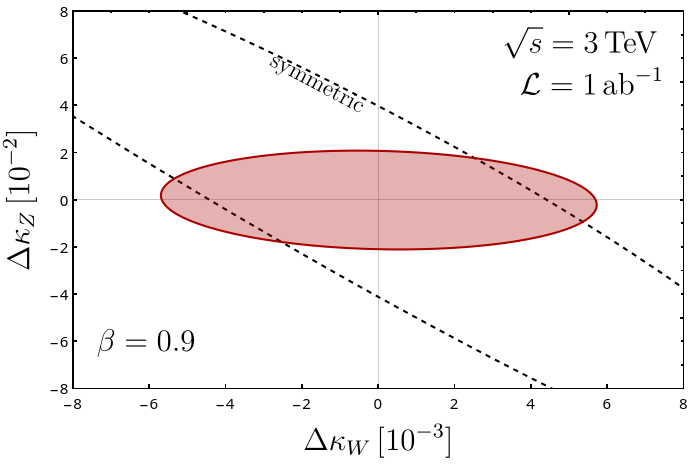}
\includegraphics[width=0.32\linewidth]{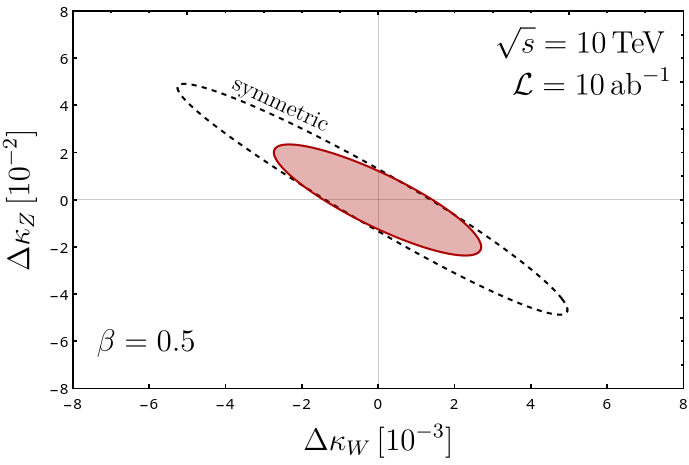}
\includegraphics[width=0.32\linewidth]{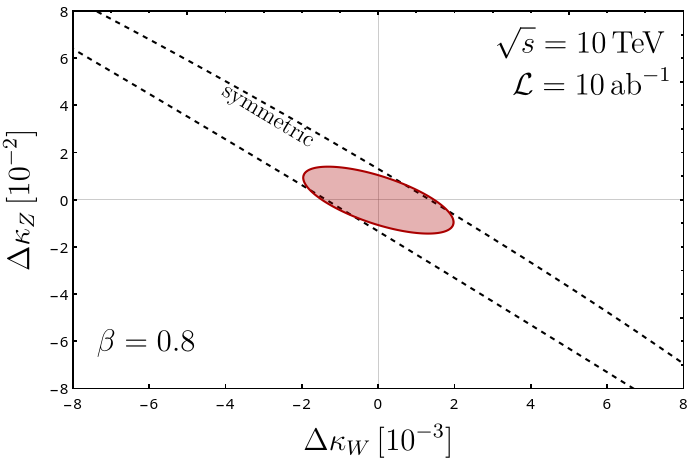}
\includegraphics[width=0.32\linewidth]{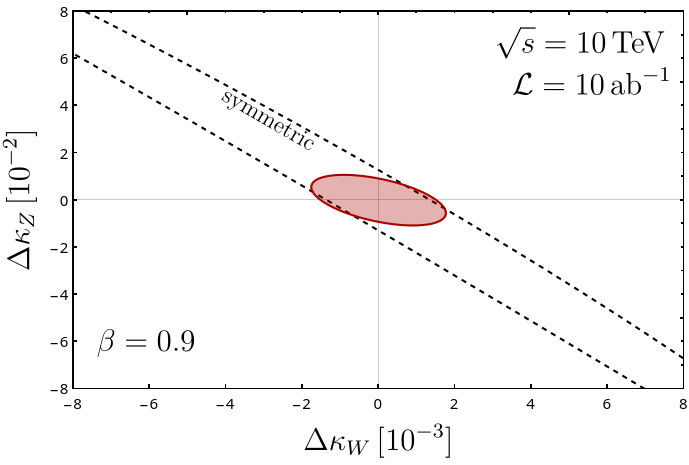}
\caption{ Combined $95\%$ C.L from the VBF Higgs process for the $\Delta \kappa_W$ vs $\Delta \kappa_Z$ for 3 TeV and 10 TeV asymmetric colliders with boosts $\beta = 0.5, 0.8, 0.9$. To compare with the asymmetric configuration, we consider a symmetric configuration with twice the energy of the most energetic beam, shown in dashed, given by (5.2, 9, 13) TeV respectively, for the 3 TeV configuration and (17, 30, 44) TeV respectively for the 10 TeV configuration. The reach for a symmetric collider with the same center of mass energy always has weaker disentangling power, and is similar to the symmetric configuration shown for $\beta =0.5$. In some cases, the symmetric higher energy collider can have slightly better reach for $\Delta \kappa_W$, but it is always worse in separating the $W$ and $Z$ initiated process. }
\label{fig:reach}
\end{figure*}

Since most of the effects of the asymmetric beam are related to only the detector geometry, we use a simplified detector simulation implemented in \texttt{DELPHES}~\cite{deFavereau:2013fsa}, using the Muon Collider Card without a forward detector. We also do not include the beam energy spread, as the goal of this work is to compare the different colliders with the same parameters. For this analysis, we do not simulate BIB, which may generate different environments in the asymmetric collider. A proper simulation of the different BIB is beyond the scope of this work.

The dominant background comes from $Z$ processes of the form
\begin{align}
    \mu^+ \mu^- &\rightarrow \bar{\nu}_{\mu} \nu_{\mu} Z(j j) \, , \\
    \mu^+ \mu^- &\rightarrow  \mu^+ \mu^- Z(j j) \, . 
\end{align}
The background can be easily handled by implementing the following basic kinematic cuts:
\begin{align}
    p_T^b &> 30 \, \text{GeV} \, , \, 10^{\circ} < \theta_{b/\mu}^{\text{lab}} < 170^{\circ} \, , \\
    m_{\text{recoil}} &> 200 \, \text{GeV} \, , \,  110 \, \text{GeV} < m_{b\bar{b}} < 140 \, \text{GeV} \, .
\end{align}

Without the inclusion of BIB effects, these cuts have sufficient distinguishing power and would not be necessary to tag the $b$-flavor, as noted in~\cite{Han:2020pif}. However, BIB effects can significantly affect the jet reconstruction as shown in Figure 4.7 of~\cite{InternationalMuonCollider:2024jyv}. For this analysis, we include b-tagging with $50\%$ efficiency to estimate the reach of such searches conservatively. 

The separate measurement of these two processes can be used to determine their respective Gauge-Higgs coupling. We can parametrize the contribution of new physics at low energies of this process using the kappa framework, where the Gauge-Higgs coupling is modified by a factor $\kappa_V = 1 + \Delta \kappa_V$. The total cross-section can be written as
\begin{align}
    \sigma_{\text{inc}} &= (1+\Delta \kappa_W)^2 \sigma_W  +  (1+\Delta \kappa_Z)^2 \sigma_Z^{\text{inc}} \, , \\
    \sigma_{1\mu} &=  (1+\Delta \kappa_Z)^2 \sigma_Z^{1\mu} \, .
\end{align}
A combined likelihood fit of $\kappa_W$ and $\kappa_Z$ is done by constructing a Poisson log-likelihood for each analysis. The $95\%$ C.L regions for the different configurations of symmetric and asymmetric colliders can be seen in Figure.~\ref{fig:reach}. The important feature of these plots is the relative behavior between the accelerator setups. There may be differences in the efficiency of the detector that change the overall size of these curves, but the relative behavior is fixed, as they are measured assuming the same detector design.

There are some interesting features that it is possible to notice from Figure.~\ref{fig:reach}. First, the effects of the asymmetry are less pronounced for lower energies. In these regions, enough muons are being detected by the tails of the distribution to obtain a good reach for both Gauge-Higgs couplings, with a slight correlation that is not possible to resolve. As the asymmetry increases, this degenerate direction gets resolved, and the search becomes statistically limited. The situation is more extreme at higher energies, where the tails of the distribution do not produce enough detectable muons, such that there is a loss of sensitivity in a direction in the $\Delta \kappa_W$ vs $\Delta \kappa_Z$ plane~\footnote{$\Delta \kappa_W \neq \Delta \kappa_Z$ breaks custodial symmetry and is a measurement of the custodial violation of the UV theory scalar potential, which need not coincide with the properties of the vacuum~\cite{deLima:2021llm,deLima:2024uwc,deLima:2024ybb} probed by Z-pole observables. }. This degeneracy gets resolved as the beam asymmetry increases.

Another thing to notice is that the symmetric configurations considered for the larger values of $\beta$ have significantly higher energy, but get significantly worse in the disentangling of the correlations between the Gauge-Higgs couplings. While the reach for $\kappa_W$ scales in these scenarios, there is effectively no reach for individual $\kappa_Z$ measurement, where all the bound comes only from the inclusive analysis.

\section{Conclusion}
\label{sec:conc}

This work proposes a different axis of optimization for muon collider designs, using asymmetric beam energies to improve forward muon reconstruction at high-energy muon colliders. In the multi-TeV regime, vector boson fusion and scattering become the dominant processes to study the electroweak sector, but their characteristic forward leptons are often lost due to shielding constraints. While ideally, forward muon detectors can access this region, there are still significant challenges to tackle on that front.

We show a partial solution, in case such detectors are not available, by introducing a modest Lorentz boost via beam asymmetry can effectively shift one of the forward muons from $Z$-fusion into the detector acceptance region, enabling their reconstruction without altering detector geometry.

Reconstructing a single forward muon is often sufficient to isolate the $Z$-fusion component of VBF, separating it from the dominant $W$-fusion background. This not only improves precision in Higgs–$Z$ coupling measurements but also enhances sensitivity to new physics in VBF topologies. 

Asymmetric $\mu e$ configurations, such as the proposed $\mu$TRISTAN~\cite{Hamada:2022mua}, offer a unique opportunity to leverage all the benefits of asymmetric beam energies. Using electrons as the low-energy beam mitigates BIB spread and improves forward coverage in the boosted direction, while the high-energy muon beam enables a sharper nozzle and extended detector reach in the opposite direction of the boost, significantly enhancing the sensitivity in forward regions. This motivates further exploration of asymmetric configurations and their experimental implications in future collider studies.

Beyond the specific case of VBF Higgs production, asymmetric beam configurations offer a general framework to recover physics in otherwise inaccessible regions of phase space. These results motivate further exploration of asymmetric beams in other forward-dominated processes, such as vector boson scattering and di-Higgs production. It is also worth exploring the effect on diverse BSM searches that have W and Z VBF topologies as backgrounds.

\acknowledgments
We thank Maximilian Ruhdorfer for the inspiring presentation in Aspen about the importance of forward muons at muon colliders. We thank David McKeen, David Morrissey, and Daniel Stolarski for helpful discussions and feedback on the draft.  This work is supported by Discovery Grants from the Natural Sciences and Engineering Research Council of Canada (NSERC). TRIUMF receives federal funding via a contribution agreement with the National Research Council (NRC) of Canada. This work was initiated in part at the Aspen Center for Physics, which is supported by a grant from the Simons Foundation (1161654, Troyer) and National Science Foundation grant PHY-2210452.

\bibliography{bibBOOST}

\end{document}